\documentclass[useAMS,usenatbib]{mn2e}
\usepackage{graphicx}
\def\jose{J. Fiestas}
\defcitealias{einsel99}{Paper~I}
\defcitealias{kim02}{Paper~II}
\defcitealias{kim04}{Paper~III}
\def\AA{Astron. \& Astroph.}

\def\AAS{Astron. \& Astrophys. Suppl.}

\def\CeMDA{Cel. Mech. Dyn. Astron.}
\def\ApJ{Astrophys. J.}

\def\ApJS{Astrophys. J. Suppl.}
\def\AJ{Astron. J.}
\def\NewA{New Astron.}

\def\PASP{Publ. Astron. Soc. Pac.}
\def\PASJ{Publ. Astron. Soc. Japan}

\def\MN{Mon. Not. Royal Astron. Soc.}

\title[2D Fokker-Planck models of Rotating clusters -- or why there is more
than King models]{2D Fokker-Planck models of Rotating clusters}
\author[\jose, R. Spurzem, E. Kim]{
\jose$^{1,2}$\thanks{E-mail:fiestas@ari.uni-heidelberg.de},
 ~R.Spurzem$^{1,2}$\thanks{E-mail:spurzem@ari.uni-heidelberg.de}
 and E. Kim $^{3}$\thanks{E-mail:ekim@head.cfa.harvard.edu}\\
$^{1}$Astronomisches Rechen-Institut, Zentrum f\"{u}r Astronomie, Universit\"{a}t Heidelberg, M\"{o}nchhofstra\ss e 12-14,   D-69120 Heidelberg, Germany\\
$^{2}$The Rhine Stellar-Dynamical Network\\
$^{3}$Harvard-Smithsonian Center for Astrophysics, 60 Garden Street, Cambridge, MA 02138.}
\begin{document}
\pagerange{\pageref{firstpage}--\pageref{lastpage}} \pubyear{2006}

\maketitle

\label{firstpage}

\begin{abstract}
Globular clusters rotate significantly, and with the increasing amount of detailed morphological and kinematical data obtained in recent years on galactic globular clusters many interesting features show up. We show how our theoretical evolutionary models of rotating clusters can be used to obtain fits, which at least properly model the overall rotation and its implied kinematics in full 2D detail (dispersions, rotation velocities). Our simplified equal mass axisymmetric rotating model provides detailed two-dimensional kinematical and morphological data for star clusters. The degree of rotation is not dominant in energy, but also non-negligible for the phase space distribution function, shape and kinematics of clusters. Therefore the models are well applicable for galactic globular clusters. Since previously published papers on that matter by us made it difficult to do detailed comparisons with observations we provide a much more comprehensive and easy-to-use set of data here, which uses as entries dynamical age and flattening of observed cluster and then offers a limited range of applicable models in full detail. The method, data structure and some exemplary comparison with observations are presented. Future work will improve modelling and data base to take a central black hole, a mass spectrum and stellar evolution into account.
\end{abstract}


\section{Introduction}

Many globular clusters do show some amount of rotation, even the old ones of our own galaxy (\citealt{meylan86,lupton87} and some ten others, K. Gebhardt, personal communication). The more accurate and star-by-star observations become available, the more important it is to properly model the effects of rotation together with evolution of globular star clusters, since typically the amount of rotational energy in clusters is not dominant, but also not negligible. It is long known that indeed flattenings of galactic globular clusters correlates with their rotation, suggesting that rotation still is important for the shape \citep{white87}. Proper Motions of stars in $\omega $ Cen have been measured to get a real 3D profile of stellar motions including rotation, and in fantastic improvements of observations using the Hubble Space Telescope it became even possible to measure rotation of 47 Tuc in the plane of the sky \citep{anderson03}. While without rotation the wealth of observational data such as luminosity functions and derived mass functions, color-magnitude diagrams, and population and kinematical analysis, obtained by e.g. the Hubble Space Telescope, for extragalactic as well as Milky Way clusters (cf. e.g. \citealt{piotto02,barmby02,pulone03,rich05,zhao05,beccari06}, to mention only the few most recently appeared papers), is balanced by a decent amount of detailed modelling (see below), theorists and observers alike seem to be appallingly puzzled by rotation in clusters, and are disappointed that (multi-mass) King model fitting does not work very well (see e.g. \citealt{piotto99b}, and for a specific example \citealt{mclaughlin03}, therein).

One notable exception being the early three-integral models by \citet{lupton87} nobody seemed to care about rotation in globular cluster modelling since then for a long time. There was a thesis by Jeremy Goodman \citeyearpar{goodman83}, whose part on rotation of clusters remained unpublished, but who was the pioneer for the main idea to use a two-dimensional approximation using the distribution function as function of energy $E$ and $z$-component of angular momentum $J_z$, and neglect the possible dependencies on a third integral $I_3$ for the time being. Unfortunately the solution of the 2D orbit averaged Fokker-Planck (henceforth FP) equation, including the self-consistent numerical determination of diffusion coefficients, is an extremely complex numerical problem, involving two levels of numerical integrations in a second order integro-differential equations, in other words we need 4D nested discretization in $E$, $J_z$, $\varpi$, and $z$ where the latter are the spatial coordinates.  It turned out that the problem was too challenging in the mid-eighties for the then available computers. With the help of Goodman, who kindly made his PhD thesis available at an early stage, \cite{einsel99} could re-write from scratch a new FP code and publish a sequence of rotating cluster models (pre-collapse, equal masses). These studies have been recently improved in a comprehensive effort on the FP side \citep{kim02,kim04}, and also promising results exist which show that the neglect of the third integral does not harm too much (if flattening is not extreme), since there is fair agreement with direct $N$-body models (see some examples in \citetalias{kim02} and \citealt{boily00}). In a new approach (\citealt{ardi05}) we are working to back up the amount of good $N$-body data for rotating clusters by a large number of small $N$ averaged models and a few large models, as was done for the non-rotating case by \cite{gierszh94,giersz96} and \cite{gierszs94}. It is clear that while all work known to the authors at this moment concentrates on self-gravitating star clusters, the improvement of our knowledge and methods in the field of rotating dense stellar systems is extremely important for galactic nuclei, too, where a central star-accreting black hole comes into the game (some stationary modelling exists, such as \citealt{quinlan90}, \citealt{murphy91}, \citealt{freitag02}). Also, claims have been made that some globular clusters in the halo of our galaxy and around M31 contain massive black holes \citep{gerssen02}, and these results have been challenged by \cite{baumgardt03a}. However, the latest paper could only provide a very limited number of case studies, due to the enormous computing time needed even on the GRAPE computers. Therefore it is very urgent to develop reliable approximate models of rotating star clusters with black hole, which is subject of current work.

Dynamical modelling of globular clusters and other collisional stellar systems (like galactic nuclei, rich open clusters, and rich galaxy clusters) still poses a considerable challenge for both theory and computational requirements (in hardware and software). On the theoretical side the validity of certain assumptions used in statistical modelling based on the FP  and other approximations is still poorly known. Stochastic noise in a discrete $N$-body system and the impossibility to directly model realistic particle numbers with the presently available hardware, are a considerable challenge for the computational side.

A large amount of individual pairwise forces between particles needs to be explicitly calculated to properly follow relaxation effects based on the cumulative effects of small angle gravitative encounters. Therefore modelling of globular star clusters over their entire lifetime requires a star-by-star simulation approach, which is highly accurate in following stellar two- and many-body encounters. Distant two--body encounters, the backbone of quasi-stationary relaxation processes, have to be treated properly by the integration method directly, while close encounters, in order to avoid truncation errors and downgrading of the overall performance, are treated in relative and specially regularized coordinates \citep{ks65,mikkola98}, their centers of masses being used in the main integrator. The factual world standard of such codes has been set by Aarseth and his codes NBODY$n$-codes ($0\leq n \leq 7$, see \citealt{aarseth99a,aarseth99b,aarseth03}). Another integrator {\sc Kira} aimed at high-precision has been used \citep{mcmillan96,portegies98,takahashi00}. It uses a hierarchical binary tree to handle compact subsystems instead of regularization.

Unfortunately, the direct simulation of most dense stellar systems with star-by-star modelling is not yet possible, although recent years have seen a significant progress in both hardware and software \citep{makino98,aarseth99a,aarseth99b,spurzem99,makino05}. Despite such progress in hardware and software even the largest useful direct $N$-body models for both globular cluster and galactic nuclei evolution have still not yet reached the realistic particle numbers ($N \sim 5\times 10^5$ for globular clusters, $N \sim 10^6$ to $10^9$ for galactic nuclei). However, recent work by \cite{baumgardt02} and \cite{baumgardt03b} has pushed the limits of present direct modelling for the first time to some $10^5$ using either NBODY6++ on parallel computers or NBODY4 on GRAPE-6 special purpose hardware. There is a notable exception of a direct one-million body problem with a central binary black hole tackled by \cite{hemsendorf03}. Their code, however, being very innovative for the large $N$ force calculation, still lacks the fine ingredients of treating close encounters between stars and black hole particles from the standard $N$-body codes.

Bridging the gap between direct models and the most interesting particle numbers in real systems is far from straightforward, neither by scaling \citep{baumgardt01}, nor by theory. There are two main classes of theory: (i) FP models, which are based on the direct numerical solution of the orbit-averaged FP equation \citep{cohn79,murphy91}, and (ii) isotropic \citep{lyndenbell80, heggie84} and anisotropic gaseous models \citep{gierszs94,spurzem96}, which can be thought of as a set of moment equations of the FP equation.

On the side of the direct FP models there have been two major recent developments. \cite{takahashi95,takahashi96,takahashi97} has published  new FP models for spherically symmetric star clusters,  based on the numerical solution of the orbit-averaged 2D FP equation (solving the FP equation for the distribution $f=f(E,J^2)$ as a function of  energy and angular momentum, on an $(E,J^2)$-mesh). \cite{drukier99} have published  results from another 2D FP code based on the original Cohn (1979) code. In such 2D FP models  anisotropy, i.e. the possible difference between radial and tangential velocity dispersions in spherical clusters, is taken into account.

Secondly, another 2D FP model has been worked out recently for the case of axisymmetric rotating star clusters \citepalias{einsel99,kim02}. Here, the distribution function is assumed to be a function of energy $E$ and the $z$-component of angular momentum $J_z$ only; a possible dependence of the distribution function on a third integral is neglected. As in the spherically symmetric case the neglection of an integral of motion is equivalent to the assumption of isotropy, here between the velocity dispersions in the meridional plane ($r$ and $z$ directions); anisotropy between velocity dispersion in the meridional plane and that in the equatorial plane ($\phi$-direction), however, is included.
We realize that the evolutionary models provided by us for rotating globular clusters are difficult to use for direct comparisons with observations, because they are not easily analytically describable. But they are the only ones which fully cope with {\it all} observational data available nowadays (full 3D velocity data, including velocity dispersions in $\varpi$ and $\phi$-direction, rotational velocity, density, all as full 2D functions of $\varpi$ and $z$). No other evolutionary model exists so far which is able to provide this information. With the advent of our new post-collapse and multi-mass models \citepalias{kim02,kim04} and the inclusion of stellar evolution and binaries (work in progress) we will be able to deliver even more interesting results. Already the existing $N$-body study \citep{ardi05} shows that rotation not only accelerates the collisional evolution but also leads to an increasing binary activity in the system.

Here we present a detailed and comprehensive set of our model data, covering a fair range of rotation rates and initial concentrations of galactic globular clusters, which is aimed to enable observers to use them for comparison with their data. The basic idea is to start up with a dynamical cluster age and a present day observed flattening, and then being able to pick a number of models from our evolutionary model data base which allow a prediction (within certain ambiguity) of how this clusters started up initially. In this paper we present the basic procedure of modelling and how we create the data, and give a few exemplary data to show the capabilities of our data base which is to be found in the web ({\tt http://www.ari.uni-heidelberg.de/clusterdata}).

Sect. 2 and 3 summarize briefly the numerical and physical procedure used for our evolving rotating cluster models, and the choice of initial conditions. Sect. 4 describes the comprehensive set of data we provide for every of our models using one example set, and Sect. 5 shows how observational data can be used to obtain the proper entry points in our data for a more detailed theory-observation comparison.

\section{Theoretical model}

The stellar system is assumed to be axisymmetric (cylindrical
coordinates $(\varpi,z,\varphi)$ were used) and in dynamical
equilibrium. Therefore, the distribution function $f(\varpi,z,\varphi)$ is almost
constant on the dynamical time scale. The only classical isolating
integrals of a general axisymmetric potential $\phi$ are the energy per unit mass:
\begin{eqnarray}
E=\frac{1}{2}v^2 + \phi(\varpi,z)
\end{eqnarray}
and the component of angular momentum along the z-axis per unit mass:
\begin{eqnarray}
J_z=\varpi v_\varphi
\end{eqnarray}
($E$, $\phi$ are negative for bound particles)\\
The Boltzmann equation written in terms of $E$ and $J_z$ has to be evaluated in the code:
\begin{eqnarray}
\frac{\partial f}{\partial t} + \frac{\partial
\phi}{\partial t}\frac{\partial f}{\partial E} = \Bigl(\frac{\partial
f}{\partial t}\Bigr)_{\rm coll}
\end{eqnarray}
$f(E,J_z,I_3)$ is the distribution function in axial symmetry. Here $I_3$ represents a third integral, which is neglected in the model (see Paper I for a discussion of possible errors).

Small angle scattering $(\Delta v/v \ll 1)$ is a condition for the expansion of the FP approximation. Diffusion coefficients of 3. and higher order are neglected. Thus, the collision term on the right side of Eq.3 is given by:
\begin{eqnarray*}
\Bigl(\frac{\partial f}{\partial t}\Bigr)_{\rm coll} = \frac{1}{V} \Bigl\lbrack-\frac{\partial}{\partial E} (\langle \Delta E \rangle f V)-\frac{\partial }{\partial J_z} (\langle \Delta J_z \rangle f V)
\end{eqnarray*}
\begin{eqnarray*}
\ \ \ \ \ \ \ \ \ \ \ \ \ \ \ \ + \frac{1}{2}  \frac{\partial^2}{\partial E^2} (\langle \Delta E^2 \rangle f V) + \frac{\partial^2}{\partial E \partial J_z}  (\langle \Delta E \Delta J_z \rangle f V)
\end{eqnarray*}
\begin{eqnarray}
\ \ \ \ \ \ \ \ \ \ \ \ \ \ \ \ + \frac{1}{2}  \frac{\partial^2}{\partial {J_z}^2}  (\langle \Delta J_z^2 \rangle f V) \Bigr\rbrack,
\end{eqnarray}
where $V$ is the volume element given by $2 \pi / \varpi$.

Due to the relation $t_{\rm r} >> t_{\rm dyn}$ (for globular clusters is $t_{\rm r} \sim 10^9 \ {\rm yr} \sim 10^3 t_{\rm dyn}$), the orbit will spread only on a relaxation timescale (because of encounters and assuming a conserved third integral). Therefore, an orbit-average of the FP equation is taken over an area in the meridional plane that intersects the hypersurface in phase space with the same $E$ and $J_z$.

Furthermore, the physical variables $E$ and $J_z$ are transformed to the suitable dimensionless quantities $X$ and $Y$. The choice of $X(E) \equiv \ln(\frac{E}{2\phi_c-E_0-E})$, where $E_0$ is the energy of a circular orbit at the core radius of the cluster, guarantees a good resolution in energy space both for small energies and for the energy belonging to the central particles. $Y(J_z,E) \equiv \frac{J_z}{J_z^{max}}$ is a dimensionless $z$-component of angular momentum ($0<|Y|<1$). A rectangular mesh is generated on which the FP equation is discretized. The models presented use a number grid size in ($X,Y$) of ($100 \times 61$). The relative error in energy conservation using this configuration is less than 0.7 per cent.

The orbit-averaged flux coefficients are derived from the local diffusion coefficients, which are found for the axial symmetric geometry via the prescriptions of \cite{rosen57} involving covariant derivatives instead of the procedure employed by Goodman (1983) using a non-covariant form. The derivation of the diffusion coefficients makes it necessary to specify a background distribution function by which test stars are scattered. Using the foreground distribution for the background self-consistently would require computational time proportional to $N_{\varpi} \times N_{\rm z} \times N_{\rm E}^2 \times N_{\rm J_z}^2$, where the $N_{\rm i}$ are grid sizes in coordinate and velocity space, which would reduce the grids employed to inaccurate descriptions of the current problem; thus, as in all previous applications concerning 2-dimensional Fokker-Planck methods for stellar systems, we set up an appropriate form of the background distribution. Herein we follow \cite{goodman83}, giving a rotating Maxwellian velocity distribution to the background, thereby reducing the computational efforts necessary.

The following steps are taken in the code\footnote [1]{This method, first developed for spherical systems in 1D by \cite{cohn79}, was originally used by \cite{goodman83} in his unpublished  PhD thesis.}:
\begin{itemize}
\item Construction of initial rotating King-models (initial potential
$\phi (\varpi,z)$ and density $n(\varpi,z)$ pairs are computed)
\item The evolution due to stellar collisions is calculated in the FP Step using the method described by \cite{henyey59}, while $\phi (\varpi,z)$ is constant
\item Recalculation of $f$ due to slow changes in the potential (VLASOV-Step).
\end{itemize}
\section{Initial conditions}

The initial models are rotating King Models of the form:
\begin{eqnarray}
f_{\rm rk} (E,J_z) = {\rm const} \times (\exp(- \beta E) -1) \times \exp(- \beta \Omega_0 J_z)
\label{king}
\end{eqnarray}
Initial parameters are:

\noindent $ \omega_0 = \sqrt{9/(4 \pi G n_{\rm c})} \cdot \Omega_0 $ dimensionless angular velocity

\noindent $ W_0 = - \beta (\phi - \phi_{\rm t}) $ dimensionless central potential,

\noindent where $ \beta = 1/{\sigma_{\rm c}}^2$, $\sigma_{\rm c}$ is central 1-dimensional velocity dispersion and $n_{\rm c}$ the central density

Time units are defined as follows:\\
Dynamical time $t_0$ \citep{cohn79}:
\begin{eqnarray}
t_0 = \sqrt{\frac{r_{\rm c_i}^3}{GM_i}} \cdot \frac{(GM_i)^2}{4 \pi \Gamma} \frac{1}{N}
\end{eqnarray}
Where, $M_i (\equiv 1)$ is the initial total cluster mass; $r_{\rm c_i} (= 1)$, the initial core radius; $N$ is the total number of particles and $\Gamma = 4 \pi G^2 (M_i/N)^2\ln \Lambda$ (ln $\Lambda$ is the Coulomb logarithm). The gravitational constant $G$ is set to 1.

The initial half mass relaxation time is defined as \citep{spitzer71}
\begin{eqnarray}
t_{\rm rh_i} = 0.138 N\sqrt {\frac {{r_{\rm h_i}}^3}{GM_i}} \frac{1}{ \ln \Lambda}
\end{eqnarray}
where $r_{\rm h_i}$ is the initial half mass radius. Following the condition $t_0 \equiv 1$, evolution of $t_{\rm rh}$ can be calculated as:
\begin{eqnarray}
t_{\rm rh} =2.208 \pi^2 \sqrt{M{r_{\rm h}}^3}
\end{eqnarray}
where $M$ and $r_{\rm h}$ are the current total mass and half-mass radius.
The cluster radius is given by
\begin{eqnarray}
r_{\rm c} \equiv \sqrt{\frac{9{\sigma_{\rm c}}^2}{4 \pi G n_{\rm c}}}
\label{corer}
\end{eqnarray}

The dynamical ellipticity $e_{\rm dyn}$ of the cluster is calculated following \cite{goodman83}, as
\begin{eqnarray}
\frac{2T_{\rm rot}+3T_{\sigma_\phi}-T_\sigma}{T_\sigma-T_{\sigma_\phi}}=\frac{(1-2s^2)\arccos{s}-3s \sqrt{1-s^2}}{s \sqrt{1-s^2}-s^2 \arccos{s}}
\label{edyn}
\end{eqnarray}
where $s \equiv b/a = 1-{\rm e_{dyn}}$, $T_{\rm rot}$ is the rotational energy, $T_{\sigma_\phi}$ is the energy contained in the azimuthal component of the velocity dispersion and $T_{\sigma}$ is the energy of all components of the velocity dispersion.
\begin{table*}
 \centering
 \begin{minipage}{140mm}
  \caption{Globular cluster observational data (see text for description)}
\begin{tabular}{|r|r|r|r|r|r|r|r|}
     \hline
     &${r_{\rm c}}$[arcmin]&${r_{\rm h}}$[arcmin]&${r_{\rm t}}$[arcmin]&$c$&e&$t_{\rm rh}$[yr]&age [yr]\\
   \hline\hline
   NGC 104 (47Tuc)&0.40&2.79&42.86&2.03&0.09&3.01$\times 10^9$&1.2$\pm$0.1$\times 10^{10}$ (4)\\
   \hline
   NGC 2808&0.26&0.76&15.55&1.77&0.12&1.3$\times 10^9$&1.6$\pm$2$\times 10^{10}$ (1)\\
   \hline
   NGC 5139 ($\omega$ Cen)&2.58 &4.80&44.83&1.24&0.12&1.0$\times 10^{10}$&1.6$\pm$0.3$\times 10^{10}$ (7)\\
   \hline
   NGC 5286&0.29&0.69&8.36&1.46&0.12&1$\times 10^9$&1.6$\pm$0.2$\times 10^{10}$ (5)\\
   \hline
   NGC 5904 (M5)&0.42&2.11&28.4&1.83&0.14&3.38$\times 10^9$&13.5$\pm$1$\times 10^9$ (6)\\
   \hline
   NGC 6093 (M80)&0.15&0.65&13.28&1.95&0.0&7.2$\times 10^8$&$1.4\times 10^{10}$ (2)\\
   \hline
   NGC 6121 (M4)&0.83&3.65&32.49&1.59&0.0&6.6$\times 10^8$&13.5$\pm$1$\times 10^9$ (6)\\
   \hline
   NGC7078 (M15)&0.07&1.06&21.50&2.5&0.05&2.23$\times 10^9$&13$\pm$3$\times 10^9$ (3)\\
   \hline\hline
\end{tabular}
\noindent {\sc References}.- (1) \citealt{alcaino90};  (2) \citealt{brocato98}; (3) \citealt{caputo84}; (4) \citealt{grundahl02};  (5) \citealt{samus95}; (6) \citealt{sandquist96}; (7) Kinematical data from \citealt{trager95}, age from \citealt{thompson01}
\label{obs}
\end{minipage}
\end{table*}

The cluster orbits its parent galaxy in a spherical potential, at a constant distance (circular orbit and constant mean density within the tidal radius). Stars are removed instantaneously beyond $r_{\rm t}$ using an energy criteria, i.e. if they have an energy larger than $E_{\rm t}(r_{\rm t})$, evaluated at every Poisson step.

\section{Numerical results}
We describe in the Appendix a set of numerical simulations classified by ellipticity $e_{\rm dyn}$ and age of the system $t/t_{\rm rh}$. The database contains following evolutionary parameters:
        \begin{itemize}
        \item  2D parameters:
                \begin{itemize}
                \item  Distribution function $f(E,Jz)$ 
                \item  Density $n(\varpi,z)$ 
                \item  1-dim velocity dispersions $\sigma_\phi(\varpi,z) $ (azimuthal), and $\sigma_{\rm r}$(=$\sigma_{\rm z}$) (in meridional plane) 
                \item  Rotational velocity $v_{\rm rot}(\varpi,z)$ and angular velocity $\Omega(\varpi,z)$ 
                \item  Potential $\phi(\varpi,z)$ and anisotropy $A(\varpi,z)$
                \end{itemize}
        \item  Global structure parameters like concentration $c$=log($r_{\rm t}/r_{\rm c}$), $(r_{\rm c}/r_{\rm h})$, escape energy and total mass \footnote [3]{See Appendix for description of data and access to database}
       \end{itemize}
The following is a sample of the database.

\begin{figure}
\includegraphics[width=85mm]{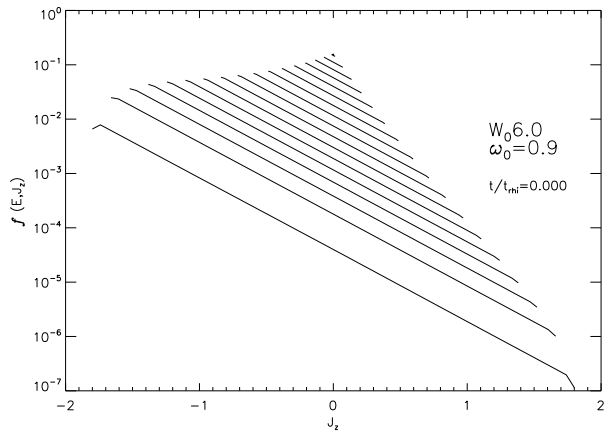}
\caption{Distribution function $f(E,J_z)$ against $J_z$ for the model ($W_0=6$, $\omega_0=0.9$) at $t=0 t_{\rm rh_i}$}
\label{df1}
\end{figure}

\begin{figure}
\includegraphics[width=85mm]{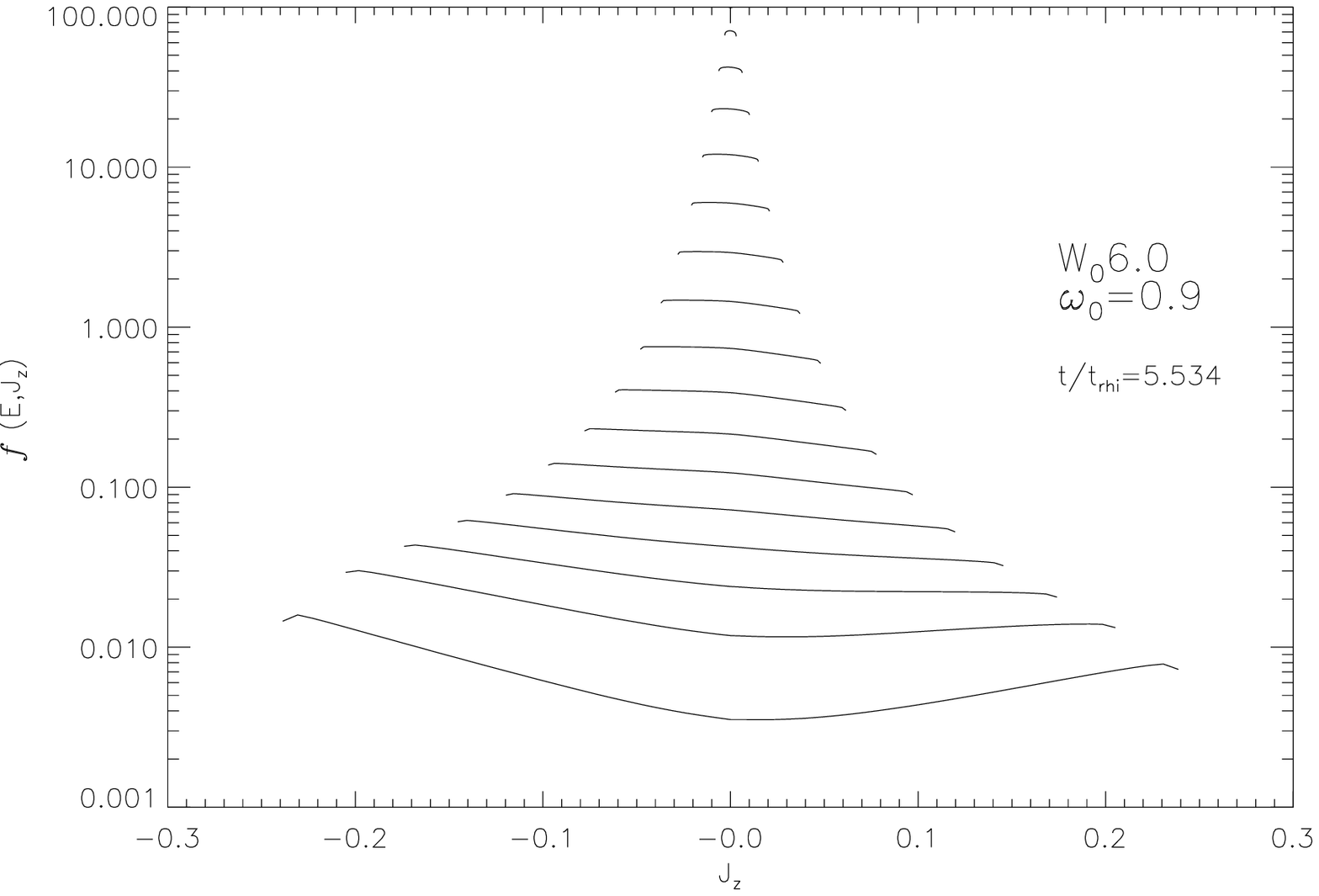}
\caption{Distribution function $f(E,J_z)$ against $J_z$ for the model ($W_0=6$, $\omega_0=0.9$) at $t=5.53 t_{\rm rh_i}$}
\label{df2}
\end{figure}

In Fig.~\ref{df1} we show the dependence of the distribution function on the angular momentum for the initial model $W_0=6$, $\omega_0=0.9$ (moderate concentration, high initial rotation) at initial time. $f(E,J_z)$ for different $J_z$ and the same energy $E$ is represented by lines. Lines of higher energy (in the core) are narrower (due to the lower values of maximal angular momentum). The slope of each curve represents the exponent in the King distribution (Eq.~\ref{king}). For comparison, we show $f(E,J_z)$ for the same model, at a later time (Fig.~\ref{df2}). We see that the slopes of the curves (rotation) change depending on the energy. They are not King-models any more, and the initial relation between $f$ and $\Omega_0$ is not conserved. Changes in anisotropy (see definition in the Appendix) can be also observed in Fig.~\ref{df2}, where orbits lower energy (bottom curves) are slightly depleted around $J_{\rm z} = 0$ (more circular orbits, negative anisotropy). At the same time one can observe positive anisotropy built by orbits of higher energy (top curves), where $f$ is enhanced around $J_{\rm z} = 0$ (more radial orbits).

Although the main source of numerical error was detected in the recalculation of $f$ due to small changes in the potential (VLASOV-step), where a second-order interpolation of $f$ is carried on; the relative error in $f$, accumulated over all grid cells per time step, is not larger than $10^{-5}$, which makes the numerical integration stable. The sharp bends at both ends of the lines of constant energy in Figs.~\ref{df1} and \ref{df2} visualize these errors. Moreover, rotating and non-rotating models show nearly the same energy conservation accumulated errors ($\sim 0.7$ per cent for a grid ($N_{\rm E}$,$N_{\rm J_z})$=(100,61)).

\begin{figure}
\includegraphics[width=85mm]{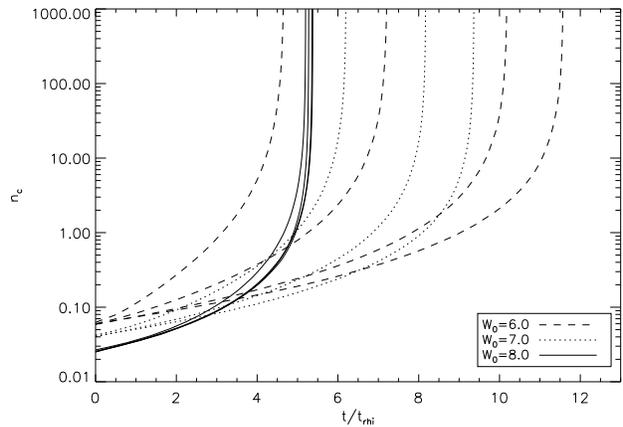}
\caption{Central density $n_{\rm c}$ against time $t/t_{\rm rh_i}$ for models with $W_0=6$ (dashed lines), $W_0=7$ (dotted lines) and  $W_0=8$ (solid lines). The rotation parameter $\omega_0$=0.0,0.3,0.5,0.7 (for $W_0=6$), $\omega_0$=0.0,0.3,0.5 (for $W_0=7$) and $\omega_0$=0.1,0.15,0.2,0.3 (for $W_0=8$) increases from right to left at the upper time axis.}
\label{density}
\end{figure}

\begin{figure}
\includegraphics[width=85mm]{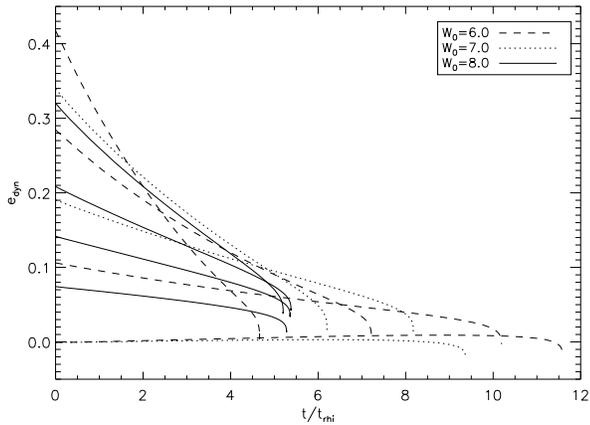}
\caption{Dynamical ellipticity $e_{\rm dyn}$ against time $t/t_{\rm rh_i}$ for models with $W_0=6$ (dashed lines), $W_0=7$ (dotted lines) and  $W_0=8$ (solid lines). The rotation parameter $\omega_0$=0.0,0.3,0.5,0.7 (for $W_0=6$), $\omega_0$=0.0,0.3,0.5 (for $W_0=7$) and $\omega_0$=0.1,0.15,0.2,0.3 (for $W_0=8$) increases from bottom to top at the $e_{\rm dyn}$ axis.}
\label{edynplot}
\end{figure}

Fig.~\ref{density} shows the evolution of the central density for 11 models with different rotation $\omega_0$ and potential $W_0$ parameters . The models with smaller parameter $\omega_0$ reach collapse in longer times. Stronger rotating models show smaller collapse times. For comparison we include here the models $W_0$=6,7,8. See \citetalias{einsel99,kim02} and \citetalias{kim04} for further discussion on collapse times. Similarly, in Fig.~\ref{edynplot} we show the evolution of dynamical ellipticity for the same models, as presented in \cite{einsel99} (their Fig. 7). As observed, ellipticity decreases steeply for the initially most strongly rotating models and the final states of all models lack significant flattening. Note that the curves can cross each other, due to less effective angular momentum transport beyond the tidal boundary, in the more moderately rotating models. In Figs.~\ref{density} and \ref{edynplot} evolution is given in units of initial $t_{\rm rh}$.
\section{Observational data comparison}
\label{obsdata}
Observations show that flattening is a common feature of globular clusters. \cite{geyer83} observed the radial variation of the ellipticity of 20 Galactic globular clusters and 4 clusters in the LMC. They found a general trend in ellipticity in a sense that in the inner and outer cluster regions they appear more circular and the ellipticity reaches a maximum at $\sim 1/3$ of the observable cluster extension. \cite{white87} have measured the projected axial ratio ($b/a$) of 100 globular clusters in the Milky Way, where $a$ and $b$ denoted the semi-major and semi-minor radii, respectively. They obtained that the mean axial ratio $<a/b>=0.93 \pm 0.01$ corresponding to $<\epsilon>=0.07 \pm 0.01$, where $\epsilon = 1- b/a$. They argued that the flattened shape of the clusters can be caused by either anisotropy in velocity dispersion or rotation.\\
\cite{meylan86} and \cite{merritt97} studied globular clusters in our galaxy to reveal without ambiguity the global rotation in clusters such as $\omega$Cen and 47Tuc. HST observations of globular clusters provide data on the velocity field of globular clusters to high accuracy \citep{vleeuwen00,anderson03}. We look forward to have direct integration models to compare with these data.\\

Table~\ref{obs} gives observational data for 6 globular clusters with ellipticities larger than zero and 2 with $e=0$. The data was taken from the Harris Catalog of Milky Way Globular Clusters \citep[][February 2003 revision]{harris96}. In the table are:

\noindent Column 1: Cluster identification number

\noindent Column 2: $r_{\rm c}$, current core radius in arc-minutes;

\noindent Column 3: $r_{\rm h}$, current half mass radius in arc-minutes;

\noindent Column 4: $r_{\rm t}$, current tidal radius in arc-minutes;

\noindent Column 5: $c \equiv {\rm log}_{10}({\rm r_t/r_c})$, concentration;

\noindent Column 6: e, ellipticity (projected axial ratio);

\noindent Column 7: $t_{\rm rh}$, half mass relaxation time in years;

\noindent Column 8: cluster age in years (taken from the literature, as remarked).

 Note that the listed values of $c$ and $r_{\rm c}$ of M15 should not be used to calculate a value of tidal radius $r_{\rm t}$ for this core-collapsed cluster as explained in the catalog.

\begin{figure}
\includegraphics[width=85mm]{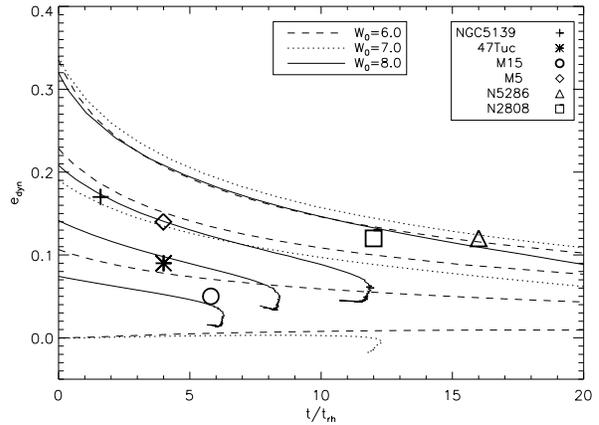}
\caption{Time evolution (in units of current $t_{\rm rh}$) of dynamical ellipticity. The rotation parameter $\omega_0$=0.0,0.3,0.5,0.7 (for $W_0=6$), $\omega_0$=0.0,0.3,0.5 (for $W_0=7$) and $\omega_0$=0.1,0.15,0.2,0.3 (for $W_0=8$) increases from bottom to top at the $e_{\rm dyn}$ axis.}
\label{ellip}
\end{figure}

A first approximation is achieved by using the age of the system in units of current half-mass relaxation time ($t/t_{\rm rh}$) and the cluster ellipticity. Fig.~\ref{ellip} shows the dependence of ellipticity on the cluster age, where the position of the identified models for six clusters of Table~\ref{obs} are included.
For the model identification, the errors in the cluster age determination (taken from the literature, Table~\ref{obs}) define age intervals, which match with the computed ellipticity curves. As commented in previous papers \citepalias{einsel99,kim02}, the observed decrease in ellipticity is not only due to angular momentum loss (evaporation) but also due to the expansion of mass shells retaining their angular momentum and decreasing their angular velocity (inversely to the actual radius of the shell). A coincidence in the behavior of ellipticity and rotational velocity profiles gives an evidence of rotation in globular clusters \citep{meylan86}.

\begin{figure}
\includegraphics[width=84mm]{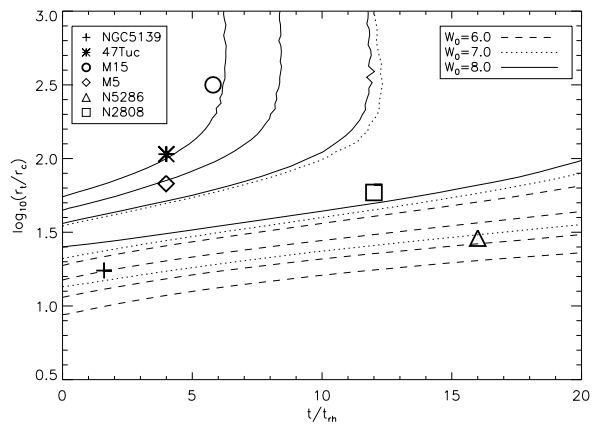}
\caption{Evolution of concentration. The rotation parameter $\omega_0$=0.0,0.3,0.5,0.7 (for $W_0=6$), $\omega_0$=0.0,0.3,0.5 (for $W_0=7$) and $\omega_0$=0.1,0.15,0.2,0.3 (for $W_0=8$) increases from top to bottom at the concentration axis.}
\label{conc}
\end{figure}

As far as concentration and ellipticity are independently computed, we use the age and concentration $c= {\rm log}(r_{\rm t}/r_{\rm c})$ of the cluster, in order to make the determination of the model more accurate. Fig.~\ref{conc}  shows the concentration against dynamical age for the same models. Using Figs.~\ref{ellip} and ~\ref{conc}, one can determine the initial parameters ($W_0$, $\omega_0$) of the model to use.

Table~\ref{models} shows the cluster parameters for GCs of Table~\ref{obs} with $e > 0$, used to determine the initial conditions of the simulation ($W_0$,$\omega_0$). First column shows the name of the cluster, second column the observed ellipticity, third column the current concentration, fourth column the age of the cluster in units of half-mass relaxation time and columns 5 and 6 the initial parameters $W_0$ (King parameter) and $\omega_0$ (initial rotation).

For the model determination the coincidence on both, ellipticity and concentration, to observations in Figs.~\ref{ellip} and \ref{conc} was located. As these figures could be filled using all possible pairs ($W_0$, $\omega_0$), we show representative initial parameters and rule out the ones which do not match both initial conditions (e.g. higher $\omega_0$, or lower $W_0$). Thus, determined models are not unique but representative. Nevertheless, as the models are still idealized, implementation of more complexity in cluster structure (different mass species, stellar evolution) and in interaction with the parent galaxy (tidal shocks, dynamical friction) could affect the results.

\begin{table}
 \centering
 \begin{minipage}{140mm}
  \caption{Globular cluster initial models (see text for description)}
\begin{tabular}{|r|r|r|r|r|r|}
     \hline
     &$e$&$c$&$t/t_{\rm rh}$&$W_0$(initial)&$\omega_0$(initial)\\
     \hline\hline
     47Tuc&0.09&2.03&4.0$\pm$0.3&8&0.15\\
     \hline
     N2808&0.12&1.77&12$\pm$1.5&8&0.3\\
     \hline
     $\omega Cen$&0.17&1.24&1.6$\pm$0.3&6&0.5\\
     \hline
     N5286&0.12&1.46&16$\pm$0.2&7&0.5\\
     \hline
     M5&0.14&1.83&3.99$\pm$0.3&8&0.2\\
     \hline
     M15&0.05&2.5&5.8$\pm$1.3&8&0.1\\
     \hline\hline
\end{tabular}
\label{models}
\end{minipage}
\end{table}

\begin{figure}
\includegraphics[width=110mm]{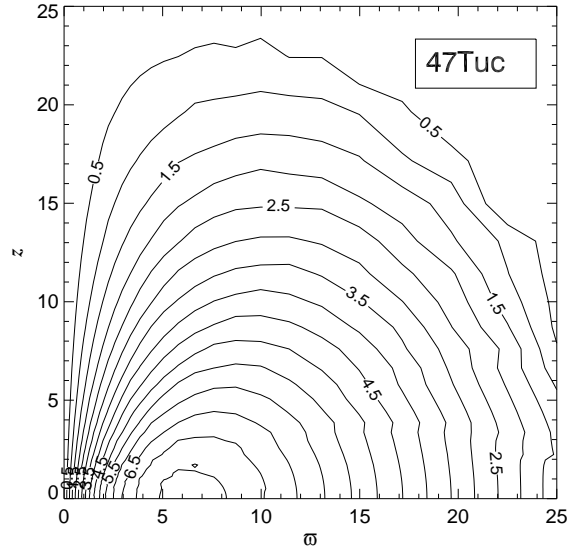}
\caption{Contour map of rotational velocity for a 47Tuc model. $W_0=8$, $\omega_0=0.15$ and $t/t_{\rm rh} = 4.0$. Distances are in arc minutes and contours are labeled in ${\rm km \ s^{-1}}$}
\label{47tuc}
\end{figure}

\begin{figure}
\includegraphics[width=110mm]{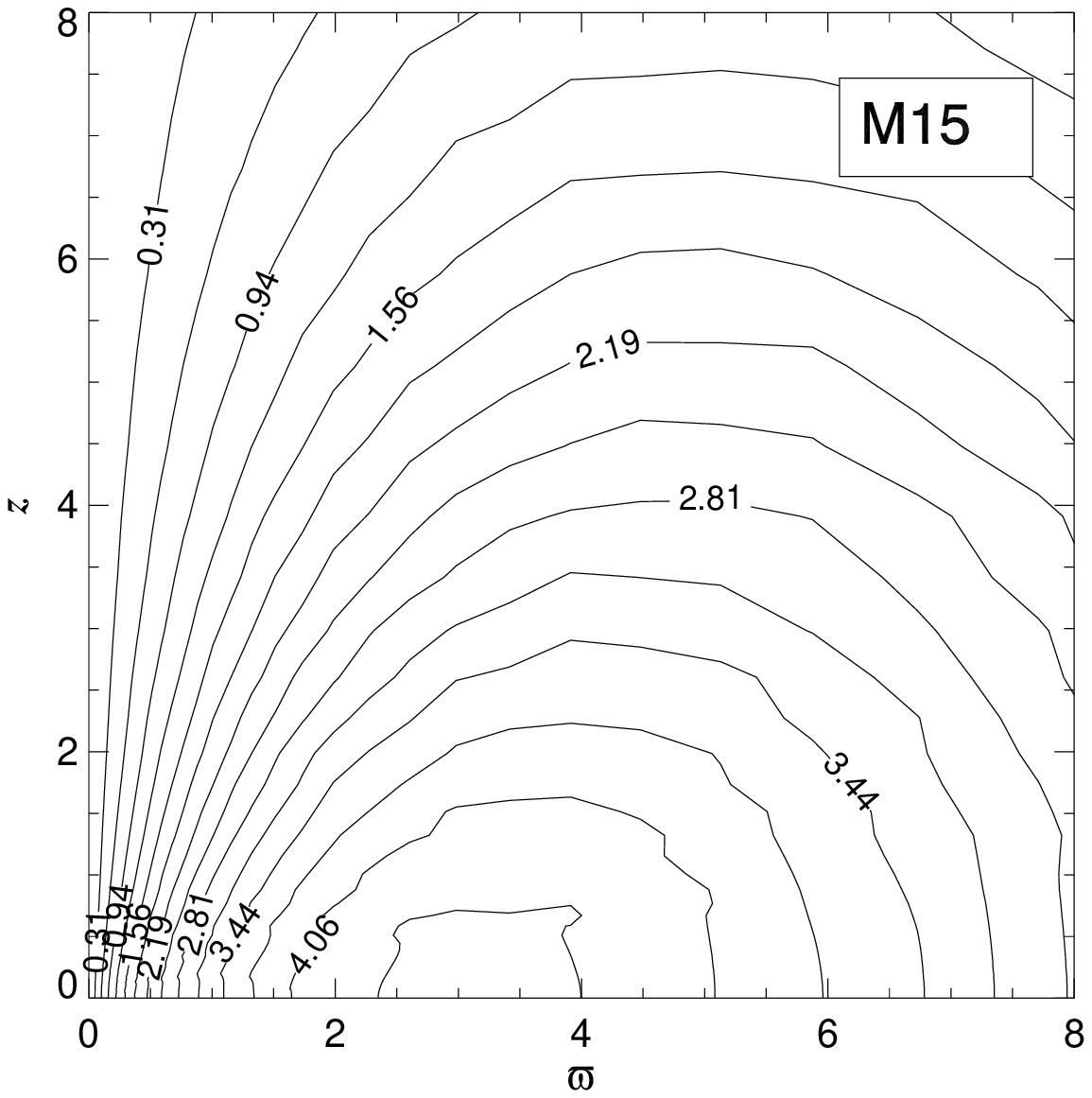}
\caption{Same as Fig.~\ref{47tuc} for a model of M15. $W_0=8$, $\omega_0=0.1$ and $t/t_{\rm rh}=5.8$}
\label{m15}
\end{figure}

\begin{figure}
\includegraphics[width=110mm]{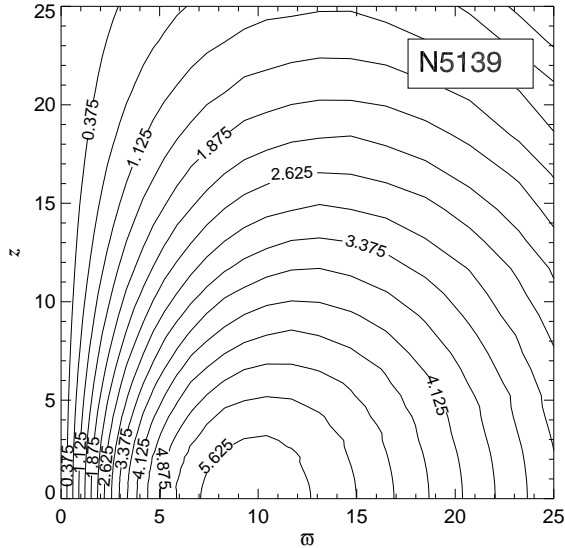}
\caption{Same as Fig.~\ref{47tuc} for a model of $\omega$Cen. $W_0=6$, $\omega_0=0.5$ and $t/t_{\rm rh}=1.6$}
\label{ocen}
\end{figure}

In Figs. \ref{47tuc}, \ref{m15}, \ref{ocen} and \ref{m5} we present contour plots of the rotational velocity $v_{\rm rot}= \varpi \ \Omega$ in the meridional plane ($\varpi,z$) for four models of Table~\ref{models}. $v_{\rm rot}$ was transformed to physical units (${\rm km/s}$) using an estimation of the total cluster mass according to a mean mass to light ratio of $M/L \ = 2$ \citep[][February 2003 revision]{harris96} and using $r_{\rm c}$ as a length scale. The distances ($\varpi$ and $z$) are rescaled to arc minutes. Single mass pre-collapsed models were used.

Fig. ~\ref{47tuc} shows a contour map of rotational velocity for an initial model ($W_0=8.0, \omega_0=0.15$) at $t/t_{\rm rh}=4$, identified as a 47 Tuc-model. The maxima of $v_{\rm rot} \sim 7.6 \ {\rm km/s}$ is located at $\sim  7.7 '$. Rotation HST measurements of 47 Tuc have recently confirmed earlier studies of this globular cluster \citep{anderson03,meylan86}. \cite{anderson03} found a rotational velocity of $5.7 \ {\rm km/s}$ at a radius of $7.5 \ {\rm pc}$. This is quite comparable to the maximum line-of-sight rotation of $\sim 6 \ {\rm km/s}$ at $5 \ {\rm pc}$ ($\sim 11-12 \ r_{\rm c}$) found by \cite{meylan86}.


\cite{gerssen02} reported the kinematical study of the central part of the globular cluster M15, including $v_{\rm rot}$. M15 is known as a globular cluster which contains a collapsed core. \cite{gebhardt94} showed that M15 has a net projected rotation amplitude of $v_{\rm rot}= 2 \ {\rm km/s}$ at radii comparable to the half-light radius (about 1'). \cite{gebhardt00} has revealed that the rotation amplitude is larger ($v_{\rm rot} = 10.4 \pm 2.7 \ {\rm km/s} $) at $r \leq 3.4''$, implying that $v_{\rm rot}/\sigma \sim 1$ in this region. Fig.~\ref{m15} shows the rotational velocity distribution in the meridional plane of a M15-model ($W_0=8.0$, $\omega_0=0.1$) at $t/t_{\rm rh}=5.8$. Although our single-mass (and multi-mass) models can not explain the rapidly rotating central region, this phenomena was already observed in our current simulations of rotating collapsed BH-models, which will be topic of a future publication. Note that the core radius of this cluster ($0.07'$) requires a better resolution to be able to confirm observational data (see below). 

\cite{gerssen02,gerssen03}, used HST to obtain new spectra of stars in the central cluster region of M15. They found theoretically, that the existence of an intermediate-mass black hole of mass $(1.7 \pm 2.7) \times 10^3M$ can explain the observed velocity dispersion. However, \cite{baumgardt03a} have shown that the core-collapse profile of a star cluster with an unseen concentration of neutron stars and heavy mass white dwarfs can explain the observed central rise of the mass-to-light ratio. Though, they did not reject the possibility of the presence of IMBHs.
\begin{figure}
\includegraphics[width=84mm]{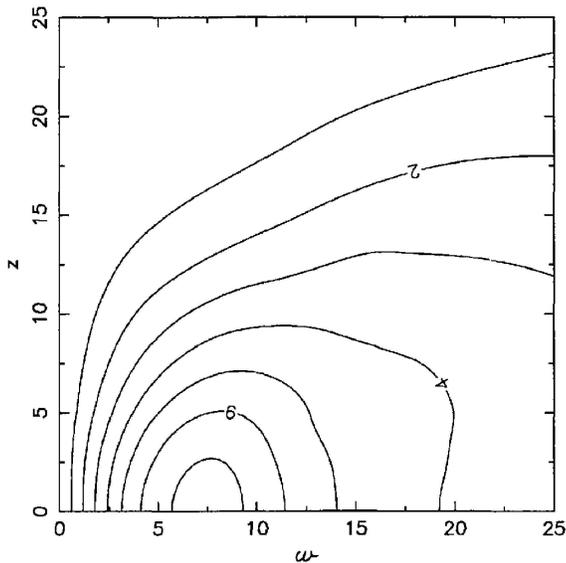}
\caption{Contour plot of the mean azimuthal velocity in the meridional plane of $\omega$Cen. Distances are in arc minutes and contours are labelled in km/s (Merritt et al. 1997).}
\label{ocenobs}
\end{figure}

Fig.~\ref{ocen} shows the contour map of rotational velocity for an initial model ($W_0=6.0, \omega_0=0.5$) at $t/t_{\rm rh}=1.6$ identified as a $\omega$Cen-model. The maxima of $v_{\rm rot}$ ($6 \ {\rm km/s}$) is found at $\sim 9' \sim 3 r_{\rm c}$. For comparison, Figure~\ref{ocenobs} shows the observed two dimensional rotational structure of $\omega$Cen in the meridional plane \citep{merritt97}. Using a non-parametric technique on 469 radial velocity data they obtained a two dimensional structure of rotational velocity in $\omega$Cen and found a peak rotational velocity of $7.9 \ {\rm km s^{-1}}$ at $\sim 11 \ {\rm pc}$, corresponding to $\sim 3$ times the core radius $(r_{\rm c})$. \cite{vleeuwen00} found agreements of proper-motion studies with the rotation in this cluster.

\begin{figure}
\includegraphics[width=110mm]{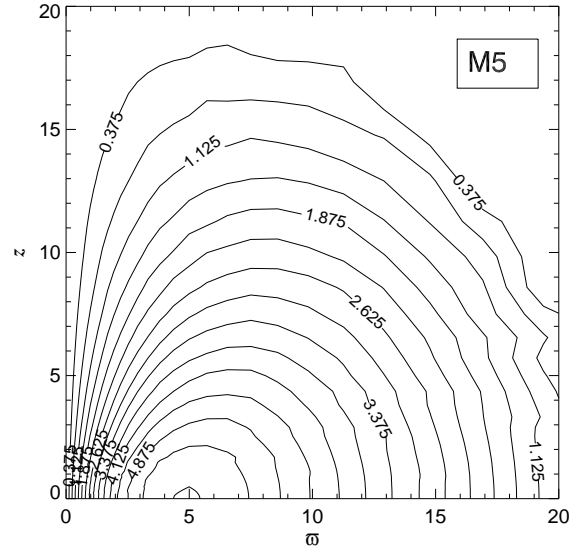}
\caption{Same as Fig.~\ref{47tuc} for a model of M5. $W_0=8$, $\omega_0=0.2$ and $t/t_{\rm rh}=4.0$}
\label{m5}
\end{figure}

Fig.~\ref{m5} shows the rotation in the meridional plane of a M5-model with a maximum of $v_{\rm rot}=5.6 \ {\rm km/sec}$ at about $5' \sim 12 r_{\rm c}$. It has one of the highest ellipticities in the Harris catalogue ($e=0.14$) and the highest in our sample. Currently there is not observational data of rotation of this globular cluster to compare with.

\begin{figure}
\includegraphics[width=85mm]{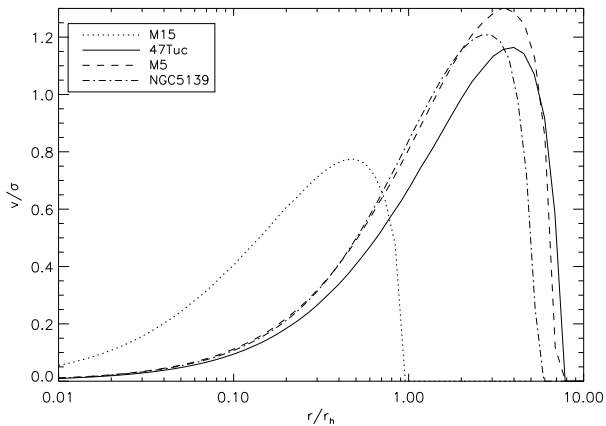}
\caption{Simulated v/$\sigma$ plotted against cluster radius, given in units of current half mass radius.}
\label{vsigma}
\end{figure}

 The simulated profile of $v_{\rm rot}/\sigma$ (rotational velocity over one dimensional velocity dispersion) of four clusters of Table~\ref{models} is plotted in Fig.~\ref{vsigma}. They show a maximum of rotation at around the half-mass radius and a rapid decline afterwards, as well as a rigid body rotation in the center. Note that the maxima of the M15-model is located closer to the center than the other clusters, as M15 has the largest dynamical age (see Table~\ref{models}) and dynamical effects as gravogyro instabilities dominate the evolution \citep{hachisu79,hachisu82}. The results show that our data is consistent with observations, as the newly presented in \cite{ven06}, who show the importance of rotation ($v_{\rm rot}/\sigma$ distribution in the meridional plane) in $\omega$Cen at radii of about 5 to 15 arcmin, however dropping more inwards and further ourwards. Nevertheless, more physical constraints (multimass modes and stellar evolution) are needed in our simulations in order to follow properly the interaction of rotation and relaxation. In the same way, we can obtain data of any model described in Section 4 and in the Appendix.  
\section{Conclusions and outlook}
Most globular clusters are flattened by rotation, and show interesting features in their rotation curves. While rotation is not dominant it is also not negligible, the total amount of rotational (ordered) kinetic energy (as compared to unordered kinetic energy) can be as high as a few 10 \% typically. Young globular clusters are usually more flattened than older ones. This was observed in the Magellanic clouds \citep{kontizas90,hill06} and in M31 \citep{barmby02}; and theoretically studied \citep{frenk82,goodwin97,bekki04}. But see also \cite{mackey05} for the MW. While observations, especially using the utmost accuracy for single star measurements in globular clusters with the Hubble Space Telescope (e.g. \citealt{anderson03,rich05}) provide a marvelous detail of morphological and sometimes surprising kinematical information, people seem sometimes still surprised that even the most elementary feature of flattening and rotation (and its dynamical implication for rotational and dispersion velocities) are so difficult to model and will not be fitted well by multi-mass King models (\citealt{longaretti97}; \citetalias{kim04}). We have provided detailed evolutionary models here which make it possible to compare them easily with observations, using data first obtained by \cite{einsel99} for the equal mass pre-collapse phase. The basic idea of our data collection, which we have discussed here in a few examples, and which is complete at the web URL {\tt http://www.ari.uni-heidelberg.de/clusterdata} is that one can go in it using simple global observational data (dynamical age, flattening and/or concentration of a cluster), and then pick from a small number of theoretical evolved models to compare with.

2-dimensional axisymmetric FP models were used in order to follow the dynamical evolution of rotating stellar systems by using initial parameters (King dimensionless central potential parameter $W_0$, dimensionless measure of rotational energy $\omega_0$). Our data sample is the only one available from theoretical evolutionary modelling of rotating globular clusters and is aimed to allow direct comparison with observational data, as the presented by \cite{merritt97} and \cite{gebhardt00}, by using our formatted data files and IDL plotting routines provided. Rotation conduces the system to a phase of strong contraction (during a mass and angular momentum loss process) and accelerates the collapse time (gravo-gyro phase; e.g. \citealt{hachisu79,hachisu82}). The observed rotational structure in the meridional plane and $v_{\rm rot}/\sigma$ curves (despite of the central region of clusters like M15, discussed in Sect.\ref{obsdata}) can be reproduced by our models. Nevertheless our models are still idealized and simplified. Physical scalings, such as time scales, may change if more effects, such of binaries, are included.

The provided database of models with theoretical cluster data will make it possible for observers to select model data to compare with \footnote [4]{(look for  http://www.ari.uni-heidelberg.de/clusterdata/)}. Our database will in the future be extended to more realism including a stellar mass spectrum and stellar evolution of singles and binaries, which all is subject of ongoing work (\citealt{ardi05}). Also a study of rotating clusters with a central massive star-accreting black hole is under way (Fiestas et al. 2006, in prep.), which will provide important data for galactic nuclei and to check claims that some globular clusters contain intermediate mass black holes.
\section*{Acknowledgments}
This work was supported in part by SFB439 (Sonderforschungsbereich) of the German Science Foundation (DFG).

\appendix

\section{Data description}
\subsection*{Grid of models}
In http://www.ari.uni-heidelberg.de/clusterdata/ we present exemplarily a grid of runs of our 2D FP models $W_0=3,6,9$ and $\omega_0=0.0$ (no-rotating), $0.3,0.6,0.9$ (rotating). The data is classified by ellipticity and age of the system. Following, a briefly description of the data.

The initial classification of the models is showed in Table 1 (on the web). ($W_0,\omega_0$) are the respectively King- and rotational initial parameters. Second table shows the grid of database. Each point of the grid represents an ellipticity and time interval. The numbers in the cells give the total number of available datasets with the corresponding age-ellipticity relation.

\begin{table*}
 \centering
 \begin{minipage}{140mm}
  \caption{}
\begin{tabular}{|r|r|}
     \hline
     File name&brief description\\
     \hline\hline
     ${\rm GP}$\_$\lbrack W_0 \times 10
\rbrack$\_$\lbrack \omega_0 \times 100 \rbrack$\_$\lbrack {\rm time
     step} \rbrack$&global parameters\\
     \hline 
    ${\rm RHO}$\_$\lbrack W_0 \times 10
\rbrack$\_$\lbrack \omega_0 \times 100 \rbrack$\_$\lbrack {\rm time step} \rbrack$&density\\
     \hline
     ${\rm ROTV}$\_$\lbrack W_0 \times 10
\rbrack$\_$\lbrack \omega_0 \times 100 \rbrack$\_$\lbrack {\rm time step} \rbrack$&rotational- and angular velocity\\
     \hline
     ${\rm VDISP}$\_$\lbrack W_0 \times 10
\rbrack$\_$\lbrack \omega_0 \times 100 \rbrack$\_$\lbrack {\rm time step} \rbrack$&1D velocity dispersion and components\\
     \hline
     ${\rm PHI}$\_${\rm AN}$\_$\lbrack W_0 \times 10
\rbrack$\_$\lbrack \omega_0 \times 100 \rbrack$\_$\lbrack {\rm time
     step} \rbrack$&potential and anisotropy\\
     \hline
     ${\rm DF}$\_$\lbrack W_0 \times 10
\rbrack$\_$\lbrack \omega_0 \times 100 \rbrack$\_$\lbrack {\rm time
     step} \rbrack$&distribution function\\
     \hline
\end{tabular}
\label{files}
\end{minipage}
\end{table*}
By selecting one grid point, one gets the datasets including following information (see Table ~\ref{files}):
\begin{enumerate}
   \item Structural parameters:\\ concentration $c={\rm log}(r_{\rm t}/r_{\rm c})$, ($r_{\rm c}/r_{\rm h}$), escape energy, total mass and energy at any time of evolution.
   \item density $n(\varpi,z)$
   \item Angular- ($\Omega(\varpi,z)$) and rotational velocity $v_{\rm rot}(\varpi,z)=\Omega(\varpi,z) \cdot \varpi$
   \item 1-dim total velocity dispersion $\sigma_{\rm t}(\varpi,z)$, azimuthal vel.disp $\sigma_\varphi (\varpi,z)$ and vel.disp. in meridional plane $\sigma_{\rm r}(\varpi,z)$
   \item Potential $\phi(\varpi,z)$ and anisotropy $A(\varpi,z)$.  
   \item Normalized (unnormalized) energy $X(E)$ and angular momentum
   $Y(J_z)$, and evaluated distribution function $f(E,J_z)$.
\end{enumerate}
Further definitions:
\begin{itemize}
   \item escape energy $x_0 \equiv \frac{\phi(r_{\rm t}) - \phi_{\rm c}}{\sigma_{\rm c}^2}$ ($\phi_{\rm c}$: central potential).
   \item core radius $r_{\rm c}$(Eq.~\ref{corer})
   \item 1-dim square velocity dispersion: $ {\sigma_{\rm t}(\varpi,z)}^2 = \sigma_\phi (\varpi,z)^2 + 2{\sigma_{\rm r}(\varpi,z)}^2$, (due to $\sigma_{\rm r}=\sigma_{\rm z}$)
   \item Anisotropy = $1-\frac{{\sigma_\phi
   (\varpi,z)}^2}{{\sigma_{\rm r}(\varpi,z)}^2}$  
   \item Normalized energy $X(E) \equiv {\rm ln}{\frac{2 \phi_{\rm c}-E-E_0}{E}}$
   \item Normalized angular momentum $Y(J_z,E) \equiv \frac{J_z}{J_z^{max}(E)}$
\end{itemize}
\subsection*{Evolution in time}
Here we present filtered data classified by initial model parameters ($W_0, \omega_0$) (Table 3 on the web). They include the time evolution of following parameters:\\\\
\begin{itemize}
    \item time in units of half mass relaxation time ($t/t_{\rm rh_i}$)
    \item time step number
    \item time in units of central relaxation time ($t/t_{\rm c}$)
    \item core radius $r_{\rm c}$ (units of initial core radius=1)
    \item central density $n_{\rm c}$
    \item central rotational velocity $v_{\rm c}$
    \item central 1-dim velocity dispersion $\sigma_c$
    \item ellipticity $e_{\rm dyn}$ (Eq.~\ref{edyn})
    \item escape energy
    \item collapse rate $\xi \equiv \frac{\tau_{rc}}{n}\frac{dn}{dt}$
    \item total mass (units of initial)
    \item total ang. momentum (z-component)
    \item total potential energy
    \item total kinetic energy
\end{itemize}
Plots as those presented in this Paper are also shown including IDL-routines used to generate them.
\bsp

\label{lastpage}


\begin{thebibliography}{99}
\bibitem[\protect\citeauthoryear{Aarseth}{1999a}]{aarseth99a} Aarseth S.J., 1999a, \PASP, 111, 1333 
\bibitem[\protect\citeauthoryear{Aarseth}{1999b}]{aarseth99b} Aarseth S.J., 1999b, \CeMDA, in press, astro-ph/9901069
\bibitem[\protect\citeauthoryear{Aarseth}{2003}]{aarseth03} Aarseth S.J., 2003, ApSS, 285, 367    
\bibitem[\protect\citeauthoryear{Alcaino et al.}{1990}]{alcaino90} Alcaino et al., 1990, \ApJS, 72, 693
\bibitem[\protect\citeauthoryear{Anderson \& King}{2003}]{anderson03} Anderson J. \& King I.R., 2003, \AJ, 126, 772
\bibitem[\protect\citeauthoryear{Ardi, Spurzem \& Mineshige}{2005}]{ardi05} Ardi E., Spurzem R. \& Mineshige S., 2005, JKAS, 38, 207
\bibitem[\protect\citeauthoryear{Barmby et al.}{2002}]{barmby02} Barmby P., 2002, \MN, 123, 1937
\bibitem[\protect\citeauthoryear{Baumgardt}{2001}]{baumgardt01} Baumgardt H., 2001, \MN, 325, 1323
\bibitem[\protect\citeauthoryear{Baumgardt, Hut \& Heggie}{2002}]{baumgardt02} Baumgardt H., Hut P. \& Heggie, D.C., 2002, \MN, 336, 1069
\bibitem[\protect\citeauthoryear{Baumgardt et al.}{2003}]{baumgardt03a} Baumgardt H. et al. 2003, \ApJ, 582, 21
\bibitem[\protect\citeauthoryear{Baumgardt \& Makino}{2003}]{baumgardt03b} Baumgardt H. \& Makino J., 2003, \MN, 340, 227
\bibitem[\protect\citeauthoryear{Beccari et al.}{2006}]{beccari06} Beccari G. et al., 2006, \AJ, 131, 2551
\bibitem[\protect\citeauthoryear{Bekki et al.}{2004}]{bekki04} Bekki K.. et al., 2004, \ApJ, 602, 730
\bibitem[\protect\citeauthoryear{Boily \& Spurzem}{2000}]{boily00} Boily C. M. \& Spurzem R., 2000, In:  The Galactic Halo, From Globular Cluster to Field Stars, Proceedings of the 35th Liege International Astrophysics Colloquium, p.607
\bibitem[\protect\citeauthoryear{Brocato et al.}{1998}]{brocato98} Brocato et al., 1998, \AA, 335, 929
\bibitem[\protect\citeauthoryear{Caputo et al.}{1984}]{caputo84} Caputo et al., 1984, \AA, 138, 457
\bibitem[\protect\citeauthoryear{Cohn}{1979}]{cohn79} Cohn H., 1979, \ApJ, 234, 1036   
\bibitem[\protect\citeauthoryear{Drukier et al.}{1999}]{drukier99} Drukier G. A. et al., 1999, \ApJ, 518, 233 
\bibitem[\protect\citeauthoryear{Einsel}{1996}]{einsel96} Einsel C., 1996, PhD Thesis, Universit\"{a}t Kiel
\bibitem[\protect\citeauthoryear{Einsel \& Spurzem}{1999, Paper~I}]{einsel99} Einsel C. \& Spurzem R., 1999, \MN, 302, 81. Paper I
\bibitem[\protect\citeauthoryear{Freitag \& Benz}{2002}]{freitag02} Freitag M. \& Benz W., 2002, \AA, 394, 345 
\bibitem[\protect\citeauthoryear{Frenk \& Fall}{1982}]{frenk82} Frenk, C.S. \& Fall, S.M., 1982, \MN, 199, 565
\bibitem[\protect\citeauthoryear{Gebhardt et al.}{1994}]{gebhardt94} Gebhardt, K. et al., 1994, \AJ 107, 2067
\bibitem[\protect\citeauthoryear{Gebhardt et al.}{2000}]{gebhardt00} Gebhardt K. et al., 2000, \ApJ, 119, 1268
\bibitem[\protect\citeauthoryear{Geyer et al.}{1983}]{geyer83} Geyer E. H., Nelles B. \& Hopp U., 1983, \AA, 125, 359
\bibitem[\protect\citeauthoryear{Gerssen et al.}{2002}]{gerssen02} Gerssen J. et al., 2002, \AJ, 124, 3270
\bibitem[\protect\citeauthoryear{Gerssen et al.}{2003}]{gerssen03} Gerssen J. et al., 2003, \AJ, 125, 376       
\bibitem[\protect\citeauthoryear{Goodman}{1983}]{goodman83} Goodman J., 1983, PhD Thesis, Princeton University
\bibitem[\protect\citeauthoryear{Giersz \& Spurzem}{1994}]{gierszs94} Giersz M. \& Spurzem R., 1994, \MN, 269, 241 
\bibitem[\protect\citeauthoryear{Giersz \& Heggie}{1994b}]{gierszh94} Giersz M. \& Heggie D.C., 1994, \MN, 270, 298
\bibitem[\protect\citeauthoryear{Giersz \& Heggie}{1996}]{giersz96} Giersz M. \& Heggie D.C., 1996, \MN, 279, 1037
\bibitem[\protect\citeauthoryear{Goodwin}{1997}]{goodwin97} Goodwin S., 1997, \MN, 286, 39
\bibitem[\protect\citeauthoryear{Grundahl et al.}{2002}]{grundahl02} Grundahl F. et al., 2002, \AA, 395, 481
\bibitem[\protect\citeauthoryear{Hachisu}{1979}]{hachisu79} Hachisu I., 1979, \PASJ, 31, 523
\bibitem[\protect\citeauthoryear{Hachisu}{1982}]{hachisu82} Hachisu I., 1982, \PASJ, 34, 313
\bibitem[\protect\citeauthoryear{Harris}{1996}]{harris96} Harris W. E., 1996, \AJ, 112, 1487. http://physun.physics.mcmaster.ca/Globular.html
\bibitem[\protect\citeauthoryear{Heggie}{1984}]{heggie84} Heggie D.C., 1984, \MN, 206, 179
\bibitem[\protect\citeauthoryear{Hemsendorf et al.}{2003}]{hemsendorf03} Hemsendorf M. et al., 2003, IAUS, 208, 405
\bibitem[\protect\citeauthoryear{Henyey et al.}{1959}]{henyey59} Henyey, L. G. et al., 1959, \ApJ, 129, 628     
\bibitem[\protect\citeauthoryear{Hill \& Zaritsky}{2006}]{hill06} Hill A. \& Zaritsky D., 2006, \AJ, 131, 414     
\bibitem[\protect\citeauthoryear{Hut}{1985}]{hut85} Hut P., 1985, in Goodman J., Hut P., eds, Proc. IAU Symp., Dynamics of Star Clusters, Reidel, Dordrecht., Vol.113, 231
\bibitem[\protect\citeauthoryear{Kim et al.}{2002 Paper~II}]{kim02} Kim E. et al., 2002, \MN, 334, 310. Paper II
\bibitem[\protect\citeauthoryear{Kim et al.}{2004 Paper~III}]{kim04} Kim E. et al., 2004, \MN, 351, 220. Paper III
\bibitem[\protect\citeauthoryear{Kontizas et al.}{1990}]{kontizas90} Kontizas, E. et al., 1990, \AJ 100, 425
\bibitem[\protect\citeauthoryear{Kustaanheimo \& Stiefel}{1965}]{ks65} Kustaanheimo P. \& Stiefel E., 1965, Reine Angew. Math., 218, 204
\bibitem[\protect\citeauthoryear{Longaretti \& Lagoute}{1997}]{longaretti97} Longaretti P.Y. \& Lagoute C., 1997, \AA, 319, 839
\bibitem[\protect\citeauthoryear{Lynden-Bell \& Eggleton}{1980}]{lyndenbell80} Lynden-Bell D. \& Eggleton P.P., 1980, \MN, 191, 483
\bibitem[\protect\citeauthoryear{Lupton, Gunn \& Griffin}{1987}]{lupton87} Lupton R.H., Gunn J.E. \& Griffin R.F. 1987, \AJ, 93, 1114
\bibitem[\protect\citeauthoryear{Mackey \& van den Berg}{2005}]{mackey05} Mackey A. \& van den Bergh S., 2005, \MN, 360, 631 
\bibitem[\protect\citeauthoryear{Makino \& Taiji}{1998}]{makino98} Makino J. \& Taiji M., 1998, Scientific simulations with special purpose computers. Chichester: Wiley, 239 pp
\bibitem[\protect\citeauthoryear{Makino}{2005}]{makino05} Makino J., 2005, JKAS, 38, 165
\bibitem[\protect\citeauthoryear{McLaughlin et al.}{2003}]{mclaughlin03} McLaughlin D. E. et al., 2003,  In: New Horizons in Globular Cluster Astronomy, ASP Conference Proceedings, Vol. 296, 101
\bibitem[\protect\citeauthoryear{McMillan \& Hut}{1996}]{mcmillan96} McMillan Stephen L. W. \& Hut P., 1996, \ApJ, 467, 348
\bibitem[\protect\citeauthoryear{Merritt et al.}{1997}]{merritt97} Merritt D. et al., 1997, \AJ, 114, 1074
\bibitem[\protect\citeauthoryear{Meylan \& Mayor}{1986}]{meylan86} Meylan G. \& Mayor M. 1986, \AA, 166, 122
\bibitem[\protect\citeauthoryear{Mikkola \& Aarseth}{1998}]{mikkola98} Mikkola S. \& Aarseth S. J., 1998, \NewA, 3, 309
\bibitem[\protect\citeauthoryear{Murphy B. et al.}{1991}]{murphy91} Murphy B., Cohn H. \& Durisen R., 1991, \ApJ, 370, 60
\bibitem[\protect\citeauthoryear{Piotto et al.}{1999}]{piotto99b} Piotto G. et al., 1999, \AJ, 118, 1727  
\bibitem[\protect\citeauthoryear{Piotto et al.}{2002}]{piotto02} Piotto G. et al., 2002, \AA, 391, 945
\bibitem[\protect\citeauthoryear{Portegies Zwart et al.}{1998}]{portegies98} Portegies Zwart S. et al., 1998, \AA, 337, 363
\bibitem[\protect\citeauthoryear{Pulone et al.}{2003}]{pulone03} Pulone L. et al., 2003, \AA, 399, 121
\bibitem[\protect\citeauthoryear{Quinlan \& Shapiro}{1990}]{quinlan90} Quinlan G. \& Shapiro S.L., 1990, \ApJ, 356, 483
\bibitem[\protect\citeauthoryear{Rich et al.}{2005}]{rich05} Rich et al., 2005, \AJ, 129, 2670
\bibitem[\protect\citeauthoryear{Rosenbluth}{1957}]{rosen57} Rosenbluth M.N., MacDonald W.M. \& Judd D.L., 1957, Physical Review, 107, 1
\bibitem[\protect\citeauthoryear{Samus et al.}{1995}]{samus95} Samus et al., 1995, \AAS, 112, 439
\bibitem[\protect\citeauthoryear{Sandquist et al.}{1996}]{sandquist96} Sandquist et al., 1996, \ApJ, 470, 910
\bibitem[\protect\citeauthoryear{Spitzer \& Hart}{1971}]{spitzer71} Spitzer, L. \& Hart M., 1971, \ApJ, 164, 399
\bibitem[\protect\citeauthoryear{Spurzem}{1996}]{spurzem96} Spurzem R., 1996. In: Dynamical Evolution of Star Clusters, IAU Symp. No. 174, 111
\bibitem[\protect\citeauthoryear{Spurzem}{1999}]{spurzem99} Spurzem R., 1999. In: Computational Astrophysics, The Journal of Computational and Applied Mathematics (JCAM), Elsevier Press, Amsterdam, in press.
\bibitem[\protect\citeauthoryear{Takahashi}{1995}]{takahashi95} Takahashi K., 1995, \PASJ, 49, 547
\bibitem[\protect\citeauthoryear{Takahashi}{1996}]{takahashi96} Takahashi K., 1996, \PASJ, 48, 691
\bibitem[\protect\citeauthoryear{Takahashi}{1997}]{takahashi97} Takahashi K., 1997, \PASJ, 47, 561
\bibitem[\protect\citeauthoryear{Takahashi \& Portegies Zwart}{2000}]{takahashi00} Takahashi K. \& Portegies Zwart S.F., 2000, \ApJ, 535, 759
\bibitem[\protect\citeauthoryear{Thompson et al.}{2001}]{thompson01} Thompson I. B. et al., 2001, \AJ, 121, 3089
\bibitem[\protect\citeauthoryear{Trager S. et al.}{1995}]{trager95} Trager S. C., King I. R. \& Djorgovski S., 1995, \AJ, 109, 218
\bibitem[\protect\citeauthoryear{Van de Ven et al.}{2006}]{ven06} Van de Ven et al., 2006, \AA, 445, 513
\bibitem[\protect\citeauthoryear{Van Leeuwen et al.}{2000}]{vleeuwen00} Van Leeuwen F. et al., 2000, \AA, 360, 472
\bibitem[\protect\citeauthoryear{White \& Shawl}{1987}]{white87} White R.E. \& Shawl S.J., 1987, \ApJ, 317, 246
\bibitem[\protect\citeauthoryear{Zhao et al.}{2005}]{zhao05} Zhao B. et al., 2005, \AJ, 129, 1934
\end{thebibliography}
\end{document}